\documentclass[12pt,english,journal,onecolumn]{IEEEtran}
\usepackage[T1]{fontenc}
\usepackage[latin9]{inputenc}
\usepackage{geometry}
\usepackage{amsmath}
\usepackage{xpatch}
\usepackage{amssymb}
\usepackage{amsfonts}
\usepackage{esint}
\usepackage{mathtools}
\usepackage{amsthm}
\usepackage{nicefrac}
\usepackage{bigints}
\usepackage{mathtools}
\usepackage{blkarray, bigstrut}
\usepackage{physics}
\usepackage{calligra}
\usepackage{graphicx}
\usepackage{epstopdf}
\usepackage{dsfont}
\usepackage{array,ragged2e}
\usepackage{enumitem}
\usepackage{etoolbox}
\usepackage{babel}
\usepackage[nopar]{lipsum}
\usepackage{psfrag}

\usepackage{balance}
\usepackage{float}
\usepackage[caption = false]{subfig}
\usepackage{setspace}
\makeatletter
\newcommand{\removelatexerror}{\let\@latex@error\@gobble}
\makeatother

\DeclareMathOperator*{\argmin}{arg\,min}

\newtheorem{theorem}{Theorem}

\newtheorem{defn}{Definition}

\xpatchcmd{\proof}{\hskip\labelsep}{\hskip5\labelsep}{}{}  
\makeatletter
\xpatchcmd{\proof}{\@addpunct{.}}{\@addpunct{:}}{}{}
\makeatother

\renewcommand\[{\begin{equation}}
\renewcommand\]{\end{equation}} 
\usepackage{listings}
\usepackage{fancyvrb}
\usepackage{framed}

\usepackage{courier}
\usepackage[usenames,dvipsnames,table]{xcolor}

\definecolor{dkgreen}{rgb}{0,0.3,0}
\definecolor{gray}{rgb}{0.5,0.5,0.5}

\DeclarePairedDelimiter\ceil{\lceil}{\rceil}

\usepackage{siunitx}
\usepackage{tabu}
\usepackage{booktabs}
\usepackage{multirow}
\usepackage[ruled,linesnumbered]{algorithm2e}

\makeatletter

\newcommand*{\rom}[1]{\expandafter\@slowromancap\romannumeral #1@}
\makeatother

\usepackage{capt-of}

\usepackage{array}
\usepackage{arydshln}
\usepackage{lscape}

\begin{document}
\title{Zero-touch Continuous Network Slicing Control via Scalable Actor-Critic Learning}
\author{Farhad Rezazadeh,~\IEEEmembership{Student Member,~IEEE}, Hatim Chergui,~\IEEEmembership{Member,~IEEE}, \\and Christos Verikoukis,~\IEEEmembership{Senior Member,~IEEE}\\
\IEEEcompsocitemizethanks{\IEEEcompsocthanksitem F. Rezazadeh, H. Chergui, and C. Verikoukis are with CTTC, Barcelona, Spain. [e-mail: \{farhad.rezazadeh, hatim.chergui, cveri\}@cttc.es].}}

\maketitle
\thispagestyle{empty}
\vspace{-1.5cm}
\begin{abstract}
Artificial intelligence (AI)-driven zero-touch network slicing is envisaged as a promising cutting-edge technology to harness the full potential of heterogeneous 5G and beyond 5G (B5G) communication systems and enable the automation of demand-aware resource management and orchestration (MANO). In this paper, we tackle the issue of B5G radio access network (RAN) joint slice admission control and resource allocation according to proposed slice-enabling cell-free massive multiple-input multiple-output (mMIMO) setup by invoking a continuous deep reinforcement learning (DRL) method. We present a novel Actor-Critic-based network slicing approach called,  \emph{prioritized twin delayed distributional deep deterministic policy gradient (D-TD3)}. The paper defines and corroborates via extensive experimental results a zero-touch network slicing scheme with a multi-objective approach where the central server learns continuously to accumulate the knowledge learned in the past to solve future problems and re-configure computing resources autonomously while minimizing latency, energy consumption, and virtual network function (VNF) instantiation cost for each slice. Moreover, we pursue a state-action return distribution learning approach with the proposed replay policy and reward-penalty mechanisms. Finally, we present numerical results to showcase the gain of the adopted multi-objective strategy and verify the performance in terms of achieved slice admission rate, latency, energy, CPU utilization, and time efficiency.
\end{abstract}

\section{Introduction}
\IEEEPARstart{N}{etwork} slicing is a paramount feature in 5G and B5G systems, wherein vertical tenants might rent and operate fully or partly isolated logical networks---or slices---on top of a physical operator's network. This is achieved by leveraging softwarization and virtualization technologies such as software-defined networking (SDN) and network functions virtualization (NFV) to yield the necessary programmability and flexibility required by network slicing to dynamically create, scale and terminate chained VNFs. The underlying resources include, e.g., physical resource blocks (PRBs), central processing unit (CPU), backhaul capacity as well as data forwarding elements (DFE) and SDN controller connections \cite{ref1}.

To ensure high throughput, reliability and energy-efficiency for multi-tenant network operation, cell-free mMIMO is the promising B5G wireless access technology to be adopted \cite{ref2}-\cite{ref3}-\cite{ref4}. It consists of a large number of distributed, low cost, low power single antenna access points (APs), connected to a network controller (i.e., the central server). The number of APs is significantly larger than the number of users \cite{ref5}. Moreover, the cell-free mMIMO exploits the favorable propagation and channel hardening properties when the number of APs is large to multiplex many users in the same time-frequency resource with small inter-user interference \cite{ref6}. This introduces new challenges on modeling cell-free networks to account for resources consumption and thereby automate the corresponding network slicing operation.

To fully automate network slicing management, ETSI has introduced the zero-touch service management (ZSM) reference framework \cite{ref7}. It is paired with the use of intent-based interfaces, closed-loop operation, and AI techniques to empower the full-automation of network slicing \cite{ref8}. This tendency towards fully automated operations and management has aroused intensive research interest in the application of AI for dynamic resource allocation, especially to tackle challenging NP-hard tasks \cite{ref9}-\cite{ref10}-\cite{ref11}. In this regard, reinforcement learning (RL) and in particular DRL methods are evaluative feedback-based schemes that can optimize and automate complex sequential decision-making tasks in B5G multi-tenant networks through the interaction with the environment and without a priori knowledge of the system. Moreover, \emph{lifelong} RL is of paramount importance because it is very hard to collect a large number of training examples in each B5G interactive environment. Alternatively, continuously learning and accumulating knowledge would yield the necessary experiences to adapt a network slicing agent to a new telecommunication environment quickly and to perform its task well \cite{ref12}. Furthermore, solving complex continuous action tasks in DRL requires a policy with stable learning, since policy may converge to sub-optimal solutions at the early stage of policy learning. In particular, the deterministic policy gradient (DPG) algorithm is a limiting case of a stochastic policy gradient \cite{ref13} in the actor-critic approach used for solving continuous tasks.

\subsection{Related Work}
Xiang \emph{et al.} have investigated the realization approach of fog RAN slicing, where two network slice instances for hotspot and vehicle-to-infrastructure scenarios are concerned and orchestrated. Moreover, they have formulated RAN slicing as an optimization problem of jointly tackling content caching and mode selection, in which the time-varying channel and unknown content popularity distribution are characterized. They proposed a DRL algorithm to maximize the reward performance under the dynamical channel state and cache status \cite{reftwc1}. However, we notice that this method cannot support scalability when the number of APs is significantly large. 

In \cite{reftwc2}, the authors have developed a novel radio slicing orchestration solution that simultaneously provides latency and throughput guarantees in a multi-tenancy environment. They have proposed a multi-armed-bandit-based (MAB) orchestrator called, LACO. It makes adaptive resource slicing decisions with no prior knowledge on the traffic demand or channel quality statistics. One of the main disadvantages of MAB is the computational complexity issue.

In \cite{ref14}, the authors have proposed a novel scheme for cell-free mMIMO networks. They have solved problem of jointly optimizing the local accuracy, transmit power, data rate, and users' processing frequency using a federated learning method. For given parameter settings, numerical results show that their joint optimization design significantly reduces the training time of federated learning over the baselines under comparison. They have also confirmed that cell-free mMIMO offers the lowest training time. However, federated learning can introduce new sources of bias through the decision of which clients to sample based on considerations such as connection type/quality, device type, location, activity patterns, and local dataset size \cite{ref15}.

Liu \emph{et al.} have studied a new decentralized DRL-based resource orchestration system, to automate dynamic network slicing in wireless edge computing networks. The DRL scheme consists of a central performance coordinator and multiple orchestration agents. They have also designed new radio, transport, and computing resource manager that enables dynamic configuration of end-to-end (E2E) resources at runtime. The evaluation results show their method achieves much improvement as compared to baseline in terms of performance, scalability, and compatibility \cite{ref16}. However, we notice that the considered state space is small.

In \cite{ref17}, the authors have developed a novel DRL algorithm for autonomous MANO of VNFs, where the central unit (CU) learns to re-configure resources (CPU and memory), to deploy new VNF instances, or to offload VNFs to a central cloud. Indeed, a deterministic policy is implemented for both action and parameter selection. They have shown that the proposed solution outperforms all benchmark DRL schemes as well as heuristic greedy allocation in a variety of network scenarios, including static traffic arrivals as well as highly time-varying traffic settings. The considered cost functions are however general and do not reflect the nature and behavior of mobile networks.

Correspondingly, \cite{ref18} has proposed a method to learn the optimum solution for demand-aware resource management in C-RAN network slicing. They have primarily proposed generative adversarial network-powered deep distributional Q-network (GAN-DDQN) and then developed dueling GAN-DDQN. Their extensive simulations have demonstrated the effectiveness of GAN-DDQN and dueling GAN-DDQN. They have considered a few metric constraints as well as stationary traffic demands pattern. This method is unable to support continuous action space in a telecommunication environment.

In \cite{ref19}, the authors have investigated the resource allocation problem in vehicular communications based on multi-agent deep deterministic policy gradient (DDPG). The goal of this work is maximizing the sum-rate of vehicle-to-infrastructure communications meanwhile guaranteeing the delivery probability of vehicle-to-vehicle communications. Their results have verified that each agent can effectively learn from the environment using the proposed DDPG algorithm for satisfying the stringent latency and reliability. Such a system with DDPG algorithm cannot guarantee to achieve robust performance because the estimation errors can lead to the agent getting stuck into a local optima or result in catastrophic forgetting.

From this state-of-the-art (SoA) overview, it turns out that to enable researchers to compare their proposed RL algorithms, there is a need to standardized RL environments for network slicing that consider network costs and reflect the behavior of B5G mobile networks. Moreover, more efforts should be deployed on RL methods that support high-dimensional state space while avoiding the curse of dimensionality, gradient explosion, and catastrophic forgetting problems. The role of continuous action approaches should also be investigated as more stable mechanisms in real environments. Besides, the design of robust and stable algorithms is considered as a necessity in B5G. 

\subsection{Contributions}
In this paper, we present the following contributions:
\begin{itemize}
    \item We propose a novel slice-enabling cell-free mMIMO access scheme with a central server while accounting for the underlying CPU, energy, and latency costs. Besides, we present well-founded models to showcase the gain of using a cross-layer design in which both baseband and transmission processing that jointly considered in the defined problem. Moreover, we investigate the feasibility of a multi-objective approach in the network slicing environment to maximize cumulative rewards while re-configuring computing resources autonomously to minimize latency, energy consumption, and VNF instantiation cost for each slice via defining correlation cost functions. This environment ensures also a reproducible comparison of different DRL algorithms through a standardized API.
    \item We adapt and fine-tune machine learning control (MLC) mechanisms on the defined environment where we adopt DRL algorithms to surmount the curse of dimensionality due to the inordinately large size of the continuous state and action spaces and thereby enable the central server learns continuously an effective control law (policy) and best actions. This is motivated by problems involving complex control tasks where it may be difficult or impossible to model the network and develop a useful control law \cite{ref21}. Furthermore, we use actor-critic architecture and pursue a joint policy and state-action return distribution optimization where we propose a prioritized asynchronous actor-learner optimized for the network slicing environment. The aim is to speed up the training process and improve learning and time efficiency. Moreover, we proposed a reward-clipping mechanism with a reward-penalty approach to mitigate the negative impact of destabilizing training. We investigate the effect of this approach on agent and learning behavior to achieve SoA performance. 
\end{itemize}
\subsection{Notation}
We summarize the main notations used throughout the paper in Table I.

\begin{table}[!htb]
\caption{Main Notations.}
\scriptsize
\centering
\begin{tabular}{@{}lc@{}}\toprule
\textbf{Notation} & \textbf{Meaning} \\ \midrule
$S$ & State Space\\ \hdashline
$A$ & Action space\\ \hdashline
$P$ & Transition probability\\ \hdashline
\textbf{$\gamma$} & Discount factor\\ \hdashline
\textbf{$\pi$} & Policy\\ \hdashline
\textbf{$V$} &  State-value function\\ \hdashline
\textbf{$Q$} & Action-value function\\ \hdashline
\textbf{$\mathcal{T}^\pi$} & Distributional Bellman operator under policy $\pi$\\ \hdashline
\textbf{$\beta$} & Experience replay\\ \hdashline
\textbf{$N$} & Number of APs\\ \hdashline
\textbf{$M$} & Number of user equipments (users)\\ \hdashline
\textbf{$L$} & number of slices\\ \hdashline
\textbf{$X$} & number of VNFs\\ \hdashline
\textbf{$\mathbf{h}_m$} & Vector of channel gains from all $N$ APs to user $m$\\ \hdashline
\textbf{$\mathbf{v}_m$} & Optimal  beamforming vector\\ \hdashline
\textbf{$(\cdot)^H$} & Conjugate transpose\\ \hdashline
\textbf{$\mathbb{C}$} & Complex set\\ \hdashline

\bottomrule
\end{tabular}

\end{table}

\subsection{Paper Outline}
The remainder of this paper is organized as follows. In Sec. \rom{2}, we describe the considered system model and the assumptions used throughout this paper and also we formulate the problem in Sec. \rom{3}. Sec. \rom{4} proposes the DRL-based network slicing resource allocation algorithm  based on a lifelong zero-touch framework. Afterwards, the performance of the proposed scheme is evaluated by simulated experiments under various settings to compare the performance of our scheme with other SoA approaches in Sec. \rom{5}. Finally, we terminate this paper with conclusion and future works in Sec. \rom{6}.

\section{System Model}

As shown in Figure 1, we consider a slice-enabling cell-free mMIMO network consisting of $N$ APs that cover $M$ single-antenna users in a downlink setup, where $N\gg M$. Let us define $L \in \mathbb{N}$ as number of slices in the network where each slice accommodates $M_l$ users with $\sum_{l=1}^{L}M_{l} = M$. We assume that all the APs are connected to central server via serial fronthaul links with sufficient capacities that runs as a set of VNFs. The central server maintains and deploys a set of VNFs to serve the users of distributed APs and also hosts agents to facilitate the training process to learn best policies and actions for scaling vertically the computing resources and consequently scale horizontally for VNFs instantiation according to system states (network configurations and parameters). For the sake of simplicity and without loss of generality, we assume 1 type of VNF per slice. We consider resource allocation tasks where the mobile network operator (MNO) collects the free and unused resources from the tenants and allocate them to the slices in need. It is done either periodically to avoid over-heading or based on requests of tenants.
\begin{figure}[ht!]
\centering
\includegraphics[scale=.36]{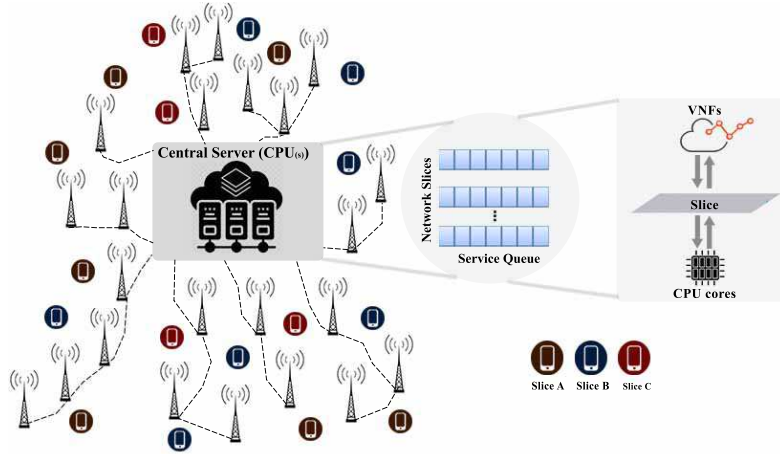}
\caption{An example of slice-enabling cell-free mMIMO network with cloud computing capability and distributed APs connected to central server via serial fronthaul links.}
\end{figure}
A total number of $X\in\mathbb{N}$ VNFs can be deployed on top of the cloud endowed with $Z\,(z=1,\ldots,Z)$ active CPUs. Each processor $z$ is characterized by a processing capability of $P_z$ million operations per time slot (MOPTS) \cite{ref30}. Let us denote $\mathbf{h}_m = [h_{1,m},h_{2,m}, ..., h_{N,M}]^H \in \mathbb{C}^{N \times 1} $ as vector of channel gains from the $N$ APs to the $M$ users, where $(\cdot)^H$ is the conjugate transpose and $\mathbb{C}$ represents the complex set. Moreover, we consider the optimal beamforming vector $\mathbf{v}_m = [v_{1,m},v_{2,m}, ..., v_{N,M}]^H \in \mathbb{C}^{N \times 1}$ associated with user $m$ and whose expression is given by \cite{ref31} as

\begin{equation}
\mathbf{v}_m = \sqrt{p_m} \frac{\left(\mathbf{I}_N + \sum_{j=1}^{M}\frac{1}{\hat{\sigma}^2}\mathbf{h}_j\mathbf{h}_j^{H}\right)^{-1}\mathbf{h}_m}{\norm{ \left(\mathbf{I}_N + \sum_{j=1}^{M}\frac{1}{\hat{\sigma}^2}\mathbf{h}_j\mathbf{h}_j^{H}\right)^{-1}\mathbf{h}_m }},  \quad m = 1, ..., M
\end{equation}
where $p_m$ is beamforming power, $\mathbf{I}_N$ denotes the $N\times N$ identity matrix and $\hat{\sigma}^2$ is the noise variance. Therefore we model the received signal $r_m \in \mathbb{C}$ at user $m$ as

\begin{equation}
r_m = \mathbf{h}_m^H \mathbf{v}_m s_m + \sum_{j \neq m}^{M}\mathbf{h}_m^H \mathbf{v}_j s_j + n_m ,\quad m = 1, ..., M
\end{equation}
where $s_j \in \mathbb{C}$ is data signal to user $m$ and received noise $n_m$ is the white Gaussian noise with zero mean and variance $\sigma^2$. Consequently, signal-to-interference-plus-noise  ratio (SINR) at user $m$ is

\begin{equation}
SINR_m = \frac{\abs{\mathbf{h}_m^H \mathbf{v}_m}^2}{\sum_{j \neq m}^{M}\abs{\mathbf{h}_m^H \mathbf{v}_j}^2+ \sigma^2 },\quad m = 1, ..., M
\end{equation}

\subsection{Computing Model}
The cloud-based baseband processing procedure at a VNF consists of several tasks such as coding, Fast Fourier Transform (FFT) and modulation. Furthermore, another computing resource is required for transmitting the signal on each AP. The corresponding computing resources follow that in \cite{ref32}. At this level let us recall the achievable rate for user $m$,

\begin{equation}
R_m = \log_2 \left( 1+ \frac{SINR_m}{\zeta_{mod}}\right),\quad m = 1, ..., M
\end{equation}
where $\zeta_{mod}$ stands for the signal-to-noise ratio (SNR) gap whit respect to modulation scheme. Therefore, the computing resources consumed by coding, FFT and modulation for user $m$ are computed by \cite{ref33}

\begin{equation}
\mathcal{C}_m^B = \hat{\theta}R_m + C_B,\quad m = 1, ..., M
\end{equation}
where $\hat{\theta}$ and $C_B$ are experimental parameters and $C_B$ is the constant complexity of the FFT function with constant computational requirements. Note that for each AP, the computing resource requirements for coding, modulation and FFT are 50\%, 10\% and 40\%, respectively \cite{ref33}. On the other hand, the signal transmission for user $m$ imposes an extra consumption of computing resource that depends on the number of non-zero elements in its network wide beamforming vector $\mathbf{v}_m$ \cite{ref30}, which is modeled as

\begin{equation}
\mathcal{C}_m^T = \delta\sum_{n = 1}^{N}\Upsilon \abs{\mathbf{v}_{n,m}},\quad m = 1, ..., M
\end{equation}
where $\delta > 0$ is slope parameter and $\Upsilon (\cdot)$  denotes step function. The whole required computing resource in the system is given by  

\begin{equation}
\mathcal{C}_{Net}^{(t)} = \sum_{m = 1}^{M}\left[\underbrace{\hat{\theta}\log_2 \left( 1+ \frac{SINR_m}{\zeta_{mod}}\right)}_{\hbox{baseband}}+\underbrace{\delta\sum_{n = 1}^{N}\Upsilon\abs{\mathbf{v}_{n,m}}}_{\hbox{transmission}}\right]+C_0
\end{equation}
where $C_0 = MC_B$ is a constant.

\subsection{Power Consumption Model}
We consider energy consumption incurred by the running processors with active VNFs and wireless transmission. We assume $\Delta_m$ is a fraction of a CPU core
\begin{equation}
\Delta_m =\hat{\theta}\log_2 \left( 1+ \frac{SINR_m}{\zeta_{mod}}\right)+\delta\sum_{n = 1}^{N}\Upsilon \abs{\mathbf{v}_{n,m}} +C_B
\end{equation}
Then the number of active CPU core is
\begin{equation}
\xi  = \ceil*{\frac{\sum_{m=1}^{M} \Delta_m}{\varsigma}},
\end{equation}
where $\varsigma$ is the capacity of a core. Each VNF is initially defined to contain $u$ cpu cores. Therefore, we have
\begin{equation}
X=\ceil*{\frac{\xi}{u}}.
\end{equation}
Thus, the total energy consumed by all active processors in Watts is given by
\begin{equation}
\mathcal{E}_{p} = \sum_{z=1}^{Z}\iota P_{z}^{3}+\sum_{x=1}^{X} \psi_{x}
\end{equation}
where $\iota$ parameter denotes the processor structure \cite{ref34} and $\psi_{x}$ is a constant value. Since the transmitted signals from APs to users have unit variance, the radio frequency power consumption at the $n-th$ AP depends only on the beamformer transmitting to the users \cite{ref35}. Let us define the total power consumption of $n-th$ AP as 

\begin{equation}
\mathcal{E}_{w}^n = \sum_{m=1}^{M}\mathbf{v}_{m}^H\mathbf{G}_{n}^H\mathbf{G}_{n}\mathbf{v}_{m}+P_n^f+P_n^c
\end{equation}
where $P_n^f$ is a constant representing the energy consumption for delivering the processed signal in fronthaul and $P_n^c$ is 
circuit power consumption. We neglect the $P_n^f$ and $P_n^f$ because usually fronthaul link is optical fiber and has small energy consumption compare to the transmit power of APs. In following \cite{ref35}, we define $\mathbf{G}_{n} \in \mathbb{C}^{1 \times N}$ as
\begin{equation}
\mathbf{G}_{n} = \left[\underbrace{\hbox{0,...,0,}}_{\hbox{n-1}} 1,0,...,0\right],\quad n > 0
\end{equation}
Therefore, the whole energy consumption in network is given by

\begin{equation}
\mathcal{E}_{Net}^{(t)} =
\underbrace{\sum_{z=1}^{Z}\iota P_{z}^{3}+\sum_{x=1}^{X} \psi_{x} }_{\hbox{baseband}}+\underbrace{\sum_{n=1}^{N}\sum_{m=1}^{M}\mathbf{v}_{m}^H\mathbf{G}_{n}^H\mathbf{G}_{n}\mathbf{v}_{m}}_{\hbox{transmission}}
\end{equation}
\subsection{Delay Model}
The  queue  prevents  loss  of  packets  that could occur when the source rate is more than the service rate and it is needed to design an appropriate admission control mechanism \cite{ref36}. According to the delay incurred at the baseband processing and also the transmission behavior in the C-RAN downlink, we use a double-layer queueing model. We model each active slice $l$ in BBU pool as a queue where data of user $m$ is processed (e.g.,  encoded) by different VNFs with mean service  rate $\hat{\mu}_m$. Then, we consider another queue for transmission of processed data, where mean service rate is $\hat{c}_m$ which satisfies $\hat{c}_m \le R_m$.

We assume that the buffer size is infinite and the tasks are served within two M/M/1 queues. The queuing model, with the channel capacity as the queue's service rate, is widely used to characterize wireless communication systems \cite{ref37}. We assume that user's packet arrival process to the processing queue is a Poisson process with mean rate $\varphi_m > 0$ \cite{ref34} and the service time of each data packet in the processing queue follows an exponential distribution with mean $\frac{1}{\hat{\mu}_m}$. Then, the arrival process to the transmitting queue is the same as the one to the  processing queue \cite{ref37}-\cite{ref38}. We suppose each tenant has just one slice and thereby from queuing theory \cite{ref39} total delay $\mathcal{D}_{q}^m$ is given by $\mathcal{D}_{q}^m=\frac{1}{\hat{\mu}_m - \varphi_m}+\frac{1}{\hat{c}_m - \varphi_m}$, Finally we have
\begin{equation}
\mathcal{D}_{Net}^{(t)}=x\mathcal{D}_{X}+\sum_{m=1}^{M}\left[\frac{1}{\hat{\mu}_m - \varphi_m}+\frac{1}{\hat{c}_m - \varphi_m}\right]
\end{equation}
where $\mathcal{D}_{X}$ denotes a constant value for creating, booting up and loading new VNFs. Considering cloud processing and wireless transmission at the central server, we define the cross-layer quality of service (QoS) constraint $\partial_m$
\begin{equation}
\partial_m=\underbrace{\mathcal{D}_{X}^m +\frac{1}{\hat{\mu}_m - \varphi_m} }_{\hbox{baseband}}+\underbrace{\frac{1}{\hat{c}_m-{\varphi}_m} }_{\hbox{transmission}}\leq \hat{\eta}
\end{equation}
where $\hat{\eta }$ is a predefined QoS requirement for each user $m$. The objective is to minimize the overall network cost incurred at each decision time step and thereby the continuous DRL optimization task is given by
\begin{subequations}
\begin{alignat}{2}
&\!\min        &\qquad& \frac{1}{M^{(t)}}(\hat{\omega}_1\mathcal{C}_{Net}^{(t)}+\hat{\omega}_2\mathcal{E}_{Net}^{(t)}+\hat{\omega}_3\mathcal{D}_{Net}^{(t)})\\
&\text{subject to} &      & p_m \leq \mathcal{P}_{max},\quad m\in M,\\
&                  &      & \frac{\abs{\mathbf{h}_m^H \mathbf{v}_m}^2}{\sum_{j \neq m}^{M}\abs{\mathbf{h}_m^H \mathbf{v}_j}^2+ \sigma^2 }\geq SINR_{th,l},\quad m\in M,l \in L,\\
&                  &      & \hat{\theta}\log_2 \left( 1+ \frac{SINR_m}{\zeta_{mod}}\right)+\delta\sum_{n = 1}^{N}\Upsilon \abs{\mathbf{v}_{n,m}} +C_B\leq \Delta_{th,l},\quad m \in M, l \in L,\\
&                  &      & \hat{c}_m \leq \log_2 \left( 1+ \frac{SINR_m}{\zeta_{mod}}\right),\quad m \in M,\\
&                  &      &\mathcal{D}_{X}^m +\frac{1}{\hat{\mu}_m - \varphi_m}+\frac{1}{\hat{c}_m-{\varphi}_m} \leq \hat{\eta }_{l}, \quad m \in M, l \in L.
\end{alignat}
\end{subequations}
where $\hat{\omega}_1, \hat{\omega}_2, \hat{\omega}_3 \in \mathbb{R}$ are fixed weights which can be set based on the operator preferences. In the section of numerical results, we validate the impact of these weights on the optimization task. Due to normalize and balance the network cost between heavy and low traffic periods, we consider the number of users ($M^{(t)}$) at each decision time step because the higher traffic can incur higher costs. Moreover, $\mathcal{P}_{max}$ is experimental value and $SINR_{th,l}$ and also $\Delta_{th,l}$ are predefined values.
\section{Problem Formulation}
In this section, we formulate the optimization problem (17a) as a Markov decision process (MDP). Our objective is to achieve lower total costs under user QoS, predefined threshold, and computing resource constraints. Due to define the reward setting of MDP closely related to our models, we propose a reward-penalty technique. From (Eqn. 1), it can be seen that the beamforming power $p_m$ for each user plays a key role in SINR at user $m$ and it has impact on the following ways. The SINR state will influence the computing resource consumption and also they are important for the mean service rate and thereby has effect on the total delay that is incurred in the network. This approach presents correlated models that find an optimal policy for selecting the best actions concerning beamforming power and computing resource allocation that should satisfy the main objective. Facing these dynamics, enable us to formulate the problem from an MDP perspective.

The MDP is a formal framework in order to employ diverse RL methods for learning optimum decision-making policies in fully or partially observable environments. The MDP for a single agent often is defined by a 5-tuple $(S,A,P,\gamma,R)$, consisting of a set of states $S$ (state space), a set of actions $A$ (action space), $P$ denotes the state transition probability for state $s$ and action $a$. It involves a decision agent that observes/perceives the state ${s}_t\in S$ of the system at each time step $t$ and then one of the available actions ${a}_t\in A$ can be selected and the system changes its state based on only ${A}_t$ and ${S}_t$  according to the probability distribution specified by $P$ so the state ${S}_t$ may change to the next state ${S}_{t+1}$. Therefore, a state transition probability is defined as ${P}_{ss^{\prime}} = \mathbb{P}[{S}_{t+1}= s^{\prime} |  {S}_t = s, {A}_t = a]$. The agent obtains an immediate reward with respect to $({s}_t,{a}_t,{s}_{t+1})$ by taking action ${a}_t$ in state ${s}_t$, ${R}_s = \mathbb{E}[{R}_{t+1} |  {S}_t = s,  {A}_t = a]$. The way we make decisions for what actions to do in what states is called a policy which is denoted with the symbol $\pi$. The return ${G}_t$ stands for the total discounted rewards from time step $t$ and the main goal is to maximize this return, ${G}_t = \sum_{n=0}^{\infty}\left(\gamma^{n} R_{t+n+1}\right)$, where $\gamma \in [0,1]$ is a real-valued discount factor that refers to how much we value rewards right now relatively to rewards in the future. This discount factor determines whether the agent is short-sighted $(\gamma=0)$ or far-sighted $(\gamma=1)$ for rewards because it is not certain about what would happen in the future. $R$ here denotes the reward function. The value function informs the agent how good is each state or action and how much reward to expect, and it is denoted $V({s}_t,{a}_t)$.
Another value function $Q$ which not only depends on the state $s$ but also on action $a$ as action-value function. The policies determine the relation between $Q$ and $V$. The value function given a policy and a state $s$ is equal to the expected value of the return, ${V}_{\pi(s)} = \mathbb{E}_{\pi}\left[\sum_{n=0}^{\infty}\left(\gamma^{n} R_{t+n+1} | S_t = s \right)\right]$, and action-value function is given by
\begin{equation}
{Q}_{\pi(s,a)} = \mathbb{E}_{\pi}\left[\sum_{n=0}^{\infty}\left(\gamma^{n} R_{t+n+1} | S_t = s, A_t = a \right)\right]
\end{equation}
The optimal policy in RL is the best policy for which there is no greater value function so for optimal value functions and optimal action-value function we have $\forall{s \in S,}\hspace{5mm}{V}_{\star}(s)={max}_{\pi}\{{V}_{\pi}(s)\}$ and $\forall{s \in S,}\hspace{2mm}{a \in A,}\hspace{5mm}{Q}_{\star}(s,a)={max}_{\pi}\{{Q}_{\pi}(s,a)\}$ respectively and the Bellman optimality equation for the action-value function satisfies by
$ {Q}_{\star}{(s,a)} = \sum_{s^\prime, r}P\left(s^\prime,r | s, a \right)\left[r+\gamma{\max}_{a^\prime}{Q}_{\star}{(s^\prime,a^\prime)}  \right]$. The Bellman equation can be used directly to solve the value function.  Q-Learning is an off-policy temporal difference learning (TD) control method,
\begin{equation}
{Q}_{t}{(s,a)} = {Q}_{t-1}{(s,a)} + \alpha\left[r+\gamma{\max}_{a^\prime}{Q}{(s^\prime,a^\prime)} - {Q}_{t-1}{(s,a)} \right]
\end{equation}
The concerned MPD problem is defined as follows,
\begin{itemize}
\item \textbf{State space:} The state space empowers the agent to have some information about possible network configurations and learn the best policy via iterations and interaction with telecommunication environment parameters. In our scenario, the state transits to the next state at each time step $t$ as input can be characterized by $S^{(t)} = \{S_1^{{(t)}}, S_2^{{(t)}}, S_3^{{(t)}}, S_4^{{(t)}}, S_5^{{(t)}}, S_6^{{(t)}}\}$, where $(S_1^{{(t)}})$ is the number of arrival requests for each slice corresponding to each VNF, $(S_2^{{(t)}})$ refers to computing resources allocated to each VNF, $(S_3^{{(t)}})$ is delay status with respect to latency cost for each slice, $(S_4^{{(t)}})$ shows energy status with respect to energy cost for each slice, $(S_5^{{(t)}})$ refers to number of users being served in each slice  and  $(S_6^{{(t)}})$ is number of VNF instantiations in each slice.
\item \textbf{Action space:} We consider vertical scaling action space for computing resources, where extra resources are added within the same logical unit to increase the capacity \cite{ref42}. More specifically, vertical scaling consists of either scaling up or down, i.e., increasing or decreasing the computing resources, respectively. In our case, the central server selects continuous value action according to traffic fluctuations and learns how to properly scaling up/down a VNF. Therefore, to change the allocated resources according to time step, we have
\begin{equation}
       \mathcal{A}_{CPU}^{(t)} \in \{ o  |   o \in \mathbb{R},-\mathcal{C}_{Net}^{(t)}\leq o \leq \mathcal{C}_{Z}^{(t)}   -\mathcal{C}_{Net}^{(t)}\}
\end{equation}
where $\mathcal{A}_{CPU}^{(t)}$ is vertical scaling action for CPU resources. One may note that vertical scaling is limited by the amount of free computing resources available on the physical server hosting the virtual machine \cite{ref43}. Although the focal point of this paper is not optimal transmit beamforming, we pursue an experimental approach to analyze the learning behaviour of the agent and its impact on the overall cost function where we define a continuous multi-action space aiming to assign beamforming power according to SINR constraint of each user belonging to each slice and explore utilizing multi-action relationship to improve the learning performance \cite{ref44} in telecommunication environment
\begin{equation}
       \mathcal{A}_{P}^{(t)} \in \{ o  |   o \in \mathbb{R}, 0\leq o \leq \mathcal{P}^{(t)}_{max}\}
\end{equation}
where $\mathcal{P}^{(t)}_{max}$ is an experimental value. the complete action space is given by
\begin{equation}
       \mathcal{A} \triangleq  \mathcal{A}_{CPU}^{(t)} \cup \mathcal{A}_{P}^{(t)}
\end{equation}

Note that we do not consider server selection and horizontal scaling because it is a discrete action and need another algorithm. Indeed, we just concentrate on parameter selection in a continuous mode that it results in decreasing VNF instantiation.
\item \textbf{Reward:} We propose a novel twin reward-penalty approach to define reward function within a telecommunication environment because the total network cost Eqn. 17a is a general and imprecise metric to guide the agent for learning good results. The main objective of agent is to learn best actions to increase the expected return while minimizing network costs. Let ${\chi}_{T}^{(t)}$ denote the constraints function that is given by the following piecewise function,

\begin{equation}
\footnotesize
\begin{aligned}
{\chi}_{T}^{(t)}=\left\{
\begin{array}{ll}
\mathbf{0},\quad \mathbf{if}\quad SINR_m \geq SINR_{th,l}\quad\mathbf{and}\quad
\Delta_m \leq \Delta_{th,l}\quad\mathbf{and}\quad\\ 
\quad\quad\quad\quad\hat{c}_m \leq R_m\quad\mathbf{and}\quad
\partial_m \leq \hat{\eta }_{l}\\
\mathbf{1},\quad \mathbf{otherwise}
                \end{array}
              \right.
\end{aligned}
\end{equation}
where $SINR_{th,l}$ stands for SINR threshold for slice $l$. Accordingly, we define the penalty cost as 
\begin{equation}
    \varepsilon_{m}^{(t)} = -\varrho_{m} \mathds{1}\left({\chi}_{T}^{(t)} = 1 \right)
\end{equation}
where $\varrho_{m}$ is the penalty coefficient for not fulfilling constraints and $\varrho_{m}^{SINR}>\varrho_{m}^{CPU}>\varrho_{m}^{MSR} > \varrho_{m}^{Delay}, m = 1, ..., M $. Consequently, the total return is given by
  \begin{equation}
    R^{(t)} =\frac{\frac{1}{ \frac{1}{M^{(t)}}(\hat{\omega}_1\mathcal{C}_{Net}^{(t)}+\hat{\omega}_2\mathcal{E}_{Net}^{(t)}+\hat{\omega}_3\mathcal{D}_{Net}^{(t)})} +  \sum_{m=1}^{M}\varepsilon_{m}^{(t)}}{\hat{\omega}_4}
  \end{equation}
where $\hat{\omega}_4$ is a hyperparameter that guarantees $R^{(t)} \in [-1, 1]$. This return function is used in deep neural network (DNN) training while the main goal of this work is the overall objective function (17a).
\end{itemize}

\section{Prioritized  Twin Delayed Distributional DDPG}
Deep learning is very powerful but neural networks can be unstable or lack of perfect convergence, so for using deep learning as function approximation in DRL, we use an experience replay buffer to store transitions. Initially, we should store random experience in the buffer. In other words, we store ${({s}_t,{a}_t,{r}_t,{s}_{t+1})}$ to train deep Q-Network and sample random many batches from the experience replay $\beta$ (buffer/queue) as training data. It results in better delineation of the true data distribution. We take a random batch $B$ for all transitions ${({s}_{t_{B}},{a}_{t_{B}},{r}_{t_{B}},{s}_{t_{B}+1})}$ of $\beta$. The goal is to predict Q-values close to target and reduce loss error over time $L^*= \frac{1}{2}\sum_{B}\left[R({s}_{t_{B}},{a}_{t_{B}})+\gamma{max}_{a}({Q}_{({s}_{t_{B}+1},a)}) - Q({{s}_{t_{B}},a}_{t_{B})}\right]^2$. The DQN-based method only works for control problems with a low-dimensional discrete action space so it is not possible to straightforwardly apply Q-learning to continuous action spaces because in continuous spaces finding the greedy policy requires optimization of $a_t$ at every time step; this optimization is too slow to be practical with large, unconstrained function approximators and non-trivial action spaces.

The proposed approach uses an actor-critic method aiming to combine the strong points of actor-only and critic-only methods. The critic uses an approximation architecture and simulation to learn a value function, which is then used to update the actor's policy parameters in a direction of performance improvement \cite{ref49}. The actor is policy (on-policy) taking the state as input and the output is actions while the critic (off-policy) takes states and actions concatenated together and returns the Q-value \cite{ref45}.

\begin{defn}
The (clipped) double Q-learning technique \cite{ref52} uses two critic networks to compute two distinguished estimates of the action-values. It benefits from two critic targets, where parameterized by ${\theta_1}$, ${\theta}_2$ and ${\theta}_1^{\prime}$,${\theta}_2^{\prime}$ respectively. Two more  DNNs, parameterized by $\phi$ and ${\phi}^{\prime}$ for actor network and actor target respectively.
\end{defn}
\begin{defn}The target in Q-learning depends on the model's prediction so cannot be considered as a true target. We use another target network instead of using Q-network to calculate the target. The Q-target is just a copy of the main Q-network periodically and the goal is to provide stability for the learning algorithm. In \cite{ref24}, it is explained that two policy networks aim to avoid overestimation.
\end{defn}
Unlike TD3 \cite{ref24} method, we use a distribution function of state-action returns for Q-value estimation instead of directly learn the Q-value (expected value) in continuous control setting and thereby replace the clipped double Q-learning with the distributional return learning. We view value functions as vectors in $\mathbb{R}^{S \times A}$, and the expected reward function as one such vector \cite{ref55}. Let define the Bellman optimality operator $\mathcal{T}^*$ as $
\mathcal{T}^*{Q}{(s,a)} = \sum_{s^\prime, r}P\left(s^{\prime},r | s, a \right)\left[r+\gamma{\max}_{a^\prime}{Q}(s^\prime,a^\prime)\right]$, where starting from any $Q_t(s, a)$, iteratively applying the operator $Q_{t+1}(s, a)\leftarrow\mathcal{T}^*{Q}{(s,a)} $ leads to converge $Q_t(s, a)\rightarrow Q^*(s, a)$ as $t \rightarrow \infty$. The optimization aim is to find function parameter such that $Q(s,a) \approx Q^*(s, a)$, and the optimal solution can be found by iteratively leveraging the $\mathcal{T}^*$ \cite{ref18}. The distributional Bellman operator preserves multimodality in value distributions, which can lead to more stable learning. Approximating the full distribution also mitigates the effects of learning from a nonstationary policy \cite{ref55}.

We first revisit Eqn. 18 in terms of the return obtained along the agent's trajectory of interactions with our custom environment (simulated network slicing environment) as a random variable $Z^\pi$ known as value distribution. We can model the distribution return $Z^\pi$ directly rather than considering expected state-action return $Q_\pi$. Then we have $Q_\pi(s, a) = \mathbb{E}\left[ Z^\pi(s, a) \right]$ and with analogous distributional approach for Bellman equation we have $Z^\pi(s, a)   \overset{\text{D}}{=}  r + \gamma Z^\pi(s^\prime, a^\prime)$. Therefore, a distributional Bellman operator $\mathcal{T}^\pi$ under policy $\pi$ can be derived as $\mathcal{T}^\pi Z(s, a)   \overset{\text{D}}{=}  r + \gamma Z(s^\prime, a^\prime)$ where we have $a^\prime \longleftarrow {\pi}_{\phi}{\prime}({s}^{\prime})+\epsilon,  \;\;\; \epsilon \sim clip(\mathcal{N}(0,{\sigma}), -c, c)$, $s^{\prime} \sim p(\cdot|s,a)$, $a^{\prime} \sim \pi(\cdot|s^{\prime})$ and also $A \overset{\text{D}}{=} B$ denotes $A$ and $B$ as two random variables with the same probability law. Let consider a deterministic policy $\pi_\phi (s)$ and define $\mathcal{Z}_\theta(\cdot|s, a)$ as parameterized state-action return distribution function. Moreover, we consider $\pi_{\phi^\prime} (s)$ and $\mathcal{Z}_{\theta^\prime}(\cdot|s, a)$ as two target networks, in order to stabilize learning. 

In traditional RL algorithms, the aim is to find optimal Q-function approximator by minimize the squared TD error but in distributional RL the goal is minimize a statistical distance to update return distribution, $\underset{ \mathcal{Z}}{\argmin}\;
\mathbb{E}\; \left[ dist \; \left( \mathcal{T}^\pi \mathcal{Z}_{old}(\cdot|s, a),\mathcal{Z}(\cdot|s,a) \right) \right]$, where $dist(A, B)$ denotes the distance between two distribution, which can be measured by some metrics, such as Kullback-Leibler (KL) divergence \cite{refnew2}. In fact, the return distribution can be optimized by minimizing the distribution distance between bellman updated and the current return distribution \cite{ref18}-\cite{ref40new5}.
\begin{defn}
Let define $A$ and $B$ as two distributions with the same support, then we can define their KL-divergence as $KL (A||B) = \int A(x)\log \frac{A(x)}{B(x)}dx$. By retrieving experience from $\beta$ we can train state-action return distribution to minimize the loss function under the KL-divergence measurement
\begin{subequations}
\begin{alignat}{2}
&J_{\mathcal{Z}}(\theta)         &\qquad&= \mathbb{E} \left[dist \left(\mathcal{T}^{{\pi}_{\phi^\prime}}\mathcal{Z}_{\theta^{\prime}}(\cdot|s,a), \mathcal{Z}_\theta(\cdot|s,a)  \right)  \right], \\
&\text{w.r.t} &      &  KL(A||B) = \sum_{i=1}^N A(x_i)\log \frac{A(x_i)}{B(x_i)} = \sum_{i=1}^N A(x_i) [\log A(x_i) - \log B(x_i)],\\
&                  &      &= \mathbb{E} \left[\sum P \left(\mathcal{T}^{{\pi}_{\phi^\prime}}Z(s,a)|\mathcal{T}^{{\pi}_{\phi^\prime}}\mathcal{Z}_{\theta^{\prime}}(\cdot|s,a)\right) \log \frac{P\left(\mathcal{T}^{{\pi}_{\phi^\prime}}Z(s,a)|\mathcal{T}^{{\pi}_{\phi^\prime}}\mathcal{Z}_{\theta^{\prime}}(\cdot|s,a)\right)}{P\left(\mathcal{T}^{{\pi}_{\phi^\prime}}Z(s,a)|\mathcal{Z}_{\theta}(\cdot|s,a)\right)} \right],\\
&                  &      &= v  - \mathbb{E} \left[\log P \left(\mathcal{T}^{{\pi}_{\phi^\prime}}Z( s, a) | \mathcal{Z}_\theta(\cdot|s,a)  \right)  \right]
\end{alignat}
\end{subequations}
where $v$ is a constant. The type of metric $dist$ that used to measure the distributional error has significant impact on performance.\end{defn} When $\mathcal{Z}_\theta$ is a continuous Gaussian or following some other distribution, the gradients $\nabla_\theta J_{\mathcal{Z}}(\theta)$ are susceptible to explode. Experimentally, we understand clipping tricks can guide agent to optimize solutions in DRL-based telecommunication environments and thereby to solve exploding gradients problem \cite{refnew1}, we clip $\mathcal{T}^{{\pi}_{\phi^\prime}}Z( s, a)$ to keep it close to the expectation
value $Q_\theta(s, a)$. Let define $clip[o, A, B]$ where $o$ is clipped into the range $[A, B]$. Therefore, we have
$\mathcal{T}^{{\pi}_{\phi^\prime}}Z( s, a) =  clip (\mathcal{T}^{{\pi}_{\phi^\prime}}Z( s, a), Q_\theta(s, a) - g, Q_\theta(s, a) + g)$, where $g$ is the clipping boundary and then the gradient is given by
\begin{equation}
\nabla_\theta J_{\mathcal{Z}}(\theta) = - \mathbb{E} [ \nabla_\theta \log P (\mathcal{T}_\mathcal{D}^{\pi_{\phi^\prime}}Z(s, a)) | \mathcal{Z}_\theta(\cdot|s, a)] 
\end{equation}
\begin{theorem}
 Suppose that the MDP satisfies conditions $p(s^{\prime}|s,a)$, $\pi_{\pi}(s)$, $\nabla_{\phi}\pi_{\phi}(s)$, $r(s,a)$, $\nabla_a r(s,a)$ where they are continuous in all parameters and variables. Consequently, it implies that deterministic policy gradient exists \cite{ref13}. Then the policy update gradient with respect to maximizing the expected return $\mathbb{E}[Q_{\theta}(s, \pi_{\phi}(s))]$,
\begin{equation}
\nabla_\phi J_{\pi}(\phi) = \mathbb{E}\left[ \nabla_{a}{Q}_{\theta}(s,a)\nabla_{\phi}\pi_{\phi}(s)\right]
\end{equation}
\end{theorem}

\begin{defn} The delayed policy updates and target networks refers to update the policy network less frequently than the value network to estimate the value with lower variance in order to have better policy \cite{ref45}, so we update parameters by
  \begin{subequations}
  \begin{equation}
  {\theta}^{\prime} \longleftarrow \tau {\theta} + (1 -\tau){\theta}^{\prime}
  \end{equation}
  \begin{equation}
  {\phi}^{\prime} \longleftarrow \tau {\phi} + (1 - \tau){\phi}^{\prime}
  \end{equation}
  \end{subequations}
where $\tau \leq1$ is an hyperparameter to tune the speed of updating. This creates a two-timescale algorithm, as often required for convergence in the linear setting \cite{ref24}-\cite{ref49}. For example, in Eqn (29b) the weights of actor network transferred to actor target slightly and actor target will be updated. In each iteration, it is going to close to actor network. This trick can stabilize the learning process.
\end{defn}
\begin{defn}
The target policy smoothing and noise regularisation considers smoothing the target policy. In any RL algorithm,  using deterministic policy entails the trade-off between exploitation and exploration. A concern with deterministic policies is they can overfit to narrow peaks in the value estimate. When updating the critic, a learning target using a deterministic policy is highly susceptible to inaccuracies induced by function approximation error, increasing the variance of the target. This induced variance can be reduced through regularization \cite{ref24} to be sure for the exploration of all possible continuous parameters. We add small random noise to target action to better exploration. We add Gaussian noise to the next action ${a}^{\prime}$ to prevent two large actions played and disturb the state of the environment. By adding this bias into action we fulfill better exploration that agent(s) do not get stuck in a state
  \begin{equation}
\tilde{a} \longleftarrow {\pi}_{\phi}{\prime}({s}^{\prime})+\epsilon,  \;\;\; \epsilon \sim clip(\mathcal{N}(0,\tilde{\sigma}), -c, c)
\end{equation}
where the noise $\epsilon$ is sampled from a Gaussian distribution with zero and certain standard deviation and clipped in a certain range of value between $-c$ and c to encourage exploration. To avoid the error of using the impossible value of actions, we clip the added noise to the range of possible actions \cite{ref24} [min\_action, max\_action] in implementation. 
\end{defn}

Experience replay is one of the main aspects of learning behaviors in biological systems. Replay buffer enables  D-TD3 agents to memorize and reuse past experiences to address catastrophic interference aiming to alleviate forgetting in the off-policy continuous control setting. Although learning from experiences by replay buffer has a significant role to stabilizes the training, but it is noisy and large due to the environmental randomness which can lead to unstable cumulative reward.
\begin{figure}[ht!]
\centering
\includegraphics[scale=0.5]{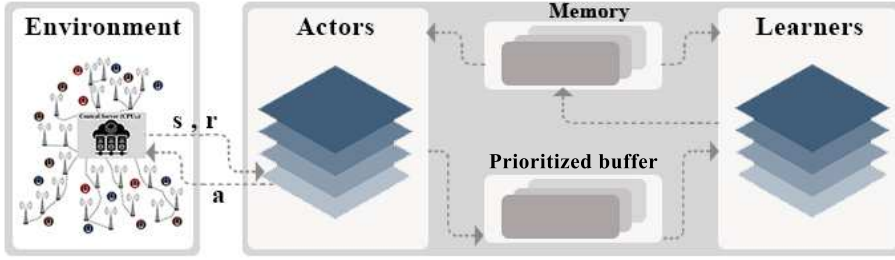}
\caption{The prioritized actor-learner architecture optimized for the network slicing  environment.}
\end{figure}
Figure 2 presents the proposed asynchronous actor-learner optimized experience replay architecture for the network slicing environment. The proposed parallelization approach enables agents to train DNN policies more stable and without large resource requirements and thereby improve time efficiency. Indeed, the parallelization approach can reduce computation time significantly while fulfilling scalability for many tasks. Unlike Gorila framework \cite{refneww4} that used separate machines to handle parameters, we use a single machine by using multithreading CPU. 

We remove the communication costs of sending parameters to other machines. Each actor with its instance of the environment generates the experiences asynchronously and sends them to buffer indiscriminately at each iteration. The stored experiences in the buffer are sent to a learner randomly as sampled data. In the next step, the learners calculate gradients locally and update the memory. The Memory consists of values and policy parameters that is used to synchronize the parameters of actors and learners. Another aim of the whole procedure is to improve sampling and updating mechanisms to fulfill better exploration and thereby obtain better overall performance and optimal policy. Instead of sample uniformly from the replay buffer, we apply a prioritized technique to distinguish the important experiences and provide the most useful samples for learners. In a standard prioritized replay approach, we use the absolute TD error to prioritize the transitions \cite{ref43new}. Due to prioritizing transitions in the distributional case, we use the minimization of KL loss. The replay buffer hosts the new transitions with maximum priority. The proposed method for network slicing is summarized in Algorithm 1.
\begin{algorithm}[ht!]
\caption{Prioritized  twin delayed distributional DDPG network slicing}
\small
\SetAlgoLined
 Initialize actor network $\phi$ and critic networks $\theta$\\
 Initialize (copy) target parameters ${\phi}^{\prime}$,  ${\theta}^{\prime}$\\
 Initialize learning rate $\ell_{\mathcal{Z}}$, $\ell_{\pi}$\\
 Initialize replay buffer $\beta$\\
 Import custom gym environment (`smartech--v1')
 
 \While {t < max\_timesteps}{
  \eIf{t < start\_timesteps}{
   $a$ = env.action\_space.sample()
   }{
   $a \longleftarrow {\pi}_{\phi}({s})+\epsilon,  \;\;\; \epsilon \sim \mathcal{N}(0,{\sigma})$
   
  }
  next\_state, reward, done, \_ = env.step($a$)\\
  store the new transition ${({s}_t,{a}_t,{r}_t,{s}_{t+1})}$ into $\beta$\\
  
   \If{t $\geq$ start\_timesteps}{
   sample $N$ batch of transitions
   {\small${({s}_{t_{B}},{a}_{t_{B}},{r}_{t_{B}},{s}_{t_{B}+1})}$} \\
   {\small\#Smoothing and noise regularisation}\\
   $a^\prime \longleftarrow {\pi}_{\phi}{\prime}({s}^{\prime})+\epsilon,  \;\;\; \epsilon \sim clip(\mathcal{N}(0,\sigma), -c, c)$\\
   {\small\#Clipping distributional Bellman operator}\\
   {\scriptsize$\mathcal{T}_\mathcal{D}^{\pi_{\phi^\prime}}Z(s, a)=  clip(\mathcal{T}_\mathcal{D}^{\pi_{\phi^\prime}}Z(s, a), Q_\theta(s, a)-g, Q_\theta (s, a)+g)$}\\ 
   {\small\#Optimize parameter $\theta$}\\
   {\scriptsize$\nabla_\theta J_{\mathcal{Z}}(\theta) = - \mathbb{E} [ \nabla_\theta \log P (\mathcal{T}_\mathcal{D}^{\pi_{\phi^\prime}}Z(s, a)) | \mathcal{Z}_\theta(\cdot | s, a)]$}

   {\small\#Update return distribution}\\
   {\small${\theta} \longleftarrow{\theta} - \ell_{\mathcal{Z}}\nabla_\theta J_{\mathcal{Z}}(\theta)$}\\
   \If{t mod freq}{
   {\small\#Compute policy update gradient}\\
    $\nabla_\phi J_{\pi}(\phi) = \mathbb{E}\left[ \nabla_{a}{Q}_{\theta}(s,a)\nabla_{\phi}\pi_{\phi}(s)\right]$\\
   {\small\#Update policy}\\
   ${\phi} \longleftarrow{\phi} + \ell_{\pi} \nabla_\phi J_{\pi}(\phi)$\\
   {\small\#Update target networks}\\
${\theta}^{\prime} \longleftarrow \tau {\theta} + (1 - \tau){\theta}^{\prime}$\\
${\phi}^{\prime} \longleftarrow \tau {\phi} + (1 - \tau){\phi}^{\prime}$

}
   }
  \If{done}{
   obs, done = env.reset(), False
   }
  $t=t+1$
 }
\end{algorithm}
\section{Performance Evaluation}

\subsection{Simulation and DRL Benchmarks}
We consider a three-tenants scenario, i.e., three slices with different QoS requirements and constraints. We assign users to different slices where users' packets arrive into the network and the algorithm computes the computing requirements to allocate to the relevant VNF when a large amount of resource is available. These slices are built upon an MNO's physical infrastructures comprised of a RAN cloud server. We consider computing resources as one of the main resources for baseband processing and signal transmission and considering a single type of VNF resources in the central server. In this respect, there exists a maximum of 50 registered subscribers randomly assigned to different slices where we dedicate more subscribers to slice-B and slice-C services and less to Sice-A services and suppose Slice-A has less frequent packets compared with the others. Moreover, we suppose these subscribers almost generate standard service traffics \cite{ref18} based on 3GPP TR 36.814 \cite{ref64} and TS 22.261 \cite{ref65}. To let heuristic beamforming perform close to the optimal beamforming we consider a setting with more antennas than users \cite{ref31}. More specifically, the number of APs in our scenario is 150. The arrival rate follows a distributed homogeneous Poisson process. Moreover, we set  $\iota = 10^{-26}$ and $P_{z} = 10 ^ 9$ \cite{ref34}, as well as $(\hat{\eta}_{l_A}, \hat{\eta}_{l_B}, \hat{\eta}_{l_C}) =(30, 65, 70)$ and $(\hat{\omega}_1, \hat{\omega}_2, \hat{\omega}_3, \hat{\omega}_4) = (1, 2, 1, 100)$.

To evaluate the quality of D-TD3, we compare our algorithm against other SoA DRL benchmark approaches, TD3 \cite{ref24}, DDPG \cite{ref45}, and SAC \cite{ref56} with a minor change for stochastic policy to keep all algorithms consistent (used clipped double Q-learning technique). Table \rom{2} provides a comparison of hyperparameters and architectures of our proposed method with other approaches. We have set the parameters following extensive experiments \cite{refglob2020}. Moreover, we use a PyTorch custom environment interfaced through OpenAI Gym, followed by the experimental parameters used in the simulations. The values of parameters depend highly on capability, scenario, and technology used. OpenAI Gym is the most famous simulation environment in the DRL community. It is a toolkit for RL research that provides a standardized way of defining the interfaces for the environment and a collection of benchmark problems that can be solved using RL. The environments are versioned in a way that will ensure that results remain meaningful and reproducible as the software is updated \cite{ref57}, specifically, it takes an action as input and provides observation, reward, done and an optional info object \cite{ref58}-\cite{ref59}.

\begin{table}
\caption{Comparison of architecture and hyperparameter tuning between algorithms.}
\scriptsize
\centering
\begin{tabular}{@{}lccccccccc@{}}\toprule
\textbf{Architecture} & \textbf{DDPG} &  \textbf{SAC} & \textbf{TD3} & \textbf{our Method (D-TD3)} \\ \midrule
\textbf{Method} & Actor-Critic  & Actor-Critic  & Actor-Critic &  Actor-Critic \\ \hdashline
\textbf{Model Type} & Multilayer perceptron  & Multilayer perceptron  & Multilayer perceptron &  Multilayer perceptron\\ \hdashline

\textbf{Policy Type} & Deterministic &  Stochastic &  Deterministic & Deterministic
\\ \hdashline
\textbf{Policy Evaluation} & TD learning &  Clipped double Q-learning & Clipped double Q-learning &  State-action distribution \\ \hdashline
\textbf{No. of DNNs}& 4  & 5  & 6  & 4 \\ \hdashline
\textbf{No. of Policy DNNs}& 1  & 1  & 1 &  1 \\ \hdashline
\textbf{No. of Value DNNs}& 1  & 2  & 2  & 1 \\ \hdashline
\textbf{No. of Target DNNs}& 2  & 2  & 3 &  2 \\ \hdashline
\textbf{No. of hidden layers}& 2  & 2  & 2  & 5 \\ \hdashline
\textbf{No. of hidden units/layer}& 200  & 256 &  400/300 & 128 \\ \hdashline
\textbf{No. of Time Steps}& $2e6$ &  $2e6$ &  $2e6$ & $2e6$ \\ \hdashline
\textbf{Batch Size}& 64  & 256  & 100  & 128 \\ \hdashline
\textbf{Optimizer}& ADAM &  ADAM &  ADAM  & ADAM \\ \hdashline
\textbf{ADAM Parameters ($\beta_1, \beta_2$)}& (0.9, 0.999) &  (0.9, 0.999)  &  (0.9, 0.999)   & (0.9, 0.999) \\ \hdashline

\textbf{Nonlinearity}& ReLU  & ReLU & ReLU &  GELU \\ \hdashline
\textbf{Target Smoothing $(\tau)$}& 0.001 &  0.005  & 0.005  & 0.001 \\ \hdashline
\textbf{Exploration Noise}& $\theta,\sigma=0.15,0.2$  & None &  $\mathcal{N}(0,0.1)$  & $\mathcal{N}(0,0.1)$ \\ \hdashline
\textbf{clamped boundary ($g$)}& None  & None &  None  & 18 \\ \hdashline

\textbf{Update Interval}& None & None  & 2  & 2 \\ \hdashline
\textbf{Policy Smoothing}& None  & None  & $\epsilon \sim\ clip(\mathcal{N}(0,0.2), -0.5, 0.5)$ &  Same-TD3 \\ \hdashline
\textbf{Expected Entropy$(\mathcal{H})$}& None & -dim(Action) & None  & None \\ \hdashline
\textbf{Actor Learning Rate}& 0.0001 &  0.0001  & 0.001  & 0.001 \\ \hdashline
\textbf{Critic Learning Rate}& 0.001 & 0.0001 &  0.001 & 0.001 \\ \hdashline
\textbf{Reward Scaling}& 1.0  & 0.2  & 1.0 & 0.2 \\ \hdashline

\textbf{Discount Factor}& 0.99  & 0.99  & 0.99  & 0.99 \\ \hdashline
\textbf{Replay Buffer Type}& Simple &  Simple &  Simple &  Prioritized  \\ \hdashline
\textbf{Replay Buffer Size}& $1e6$ &  $1e6$ &  $1e6$ &  $1e6$ \\ \hdashline
\textbf{No. of (Actor, Buffer, Learner)}& None & None & None &  (3, 2, 3) \\ \hdashline
\textbf{Max Episode Length}& $50$ &  $50$ &  $50$ &  $50$ \\
\hdashline
\textbf{Seed}& System time &  System time &  System time &  System time \\
\hdashline

\bottomrule
\end{tabular}

\end{table}

The DNNs structure for the actor-critic networks and target networks are the same. The inputs are processed by a fully connected network with 5 hidden layers where each layer consisting of 128 units and using Gaussian error linear units (GELU) \cite{ref60} for both the actor and critic. The network parameters are updated using Adam \cite{ref61}. Once the buffer $\beta$ is filled with enough samples, we train networks via a mini-batch of 128 transitions to compute the losses of the actor and critic. The buffer size is 1 million and consists of the entire history of the agent and also $start\_timestep$ denotes the time steps for initial random policy. The training process is conducted through 3 different runs and with different random seeds where the maximum length of each episode is 50 time steps. Each task is run for 2 million time steps with evaluations every 20000 iterations. Meanwhile, the evaluation computes the average return over the best 3 of 5 episodes. To delay policy updates in the algorithm, we set $freq = 2$ aiming to update the policy and target networks every 2 iterations.

\subsection{Numerical Results}
In this section, we present the numerical results obtained with the D-TD3 algorithm and other benchmarks. As we mentioned in Sec. \rom{3}, the problem formulation is general so we use constraints and thresholds as a penalty in the implementation to lead the agent to the good results and this is the reason for negative values in the learning curves. The algorithm learns according to iterations or interact with the environment with different parameters of the network and thereby the results experience high fluctuation during learning.\\
\emph{\textbf{-Learning curve:}} The learning curves are depicted in Figure 3-(a) and the related results in Table \rom{3}. They demonstrate that D-TD3 outperforms all other baseline algorithms on the main metric of this work across the corresponding task. 
\begin{figure}[ht!]
\centering
\hspace*{\fill}%
\subfloat[Learning curves of the OpenAI gym network slicing environment and continuous control benchmarks. The curves are smoothed for visual clarity concerning confidence interval.]{%
      \includegraphics[scale=.7]{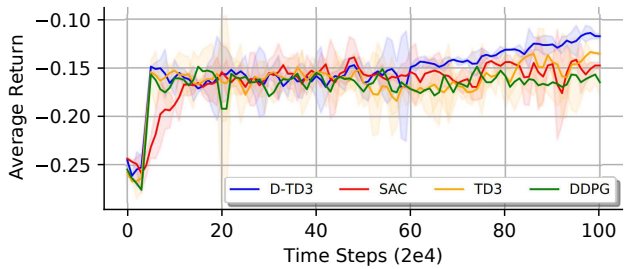}}
    \hfill
\subfloat[Time efficiency comparison of different algorithms on the custom environment.]{%
     \includegraphics[scale=.45]{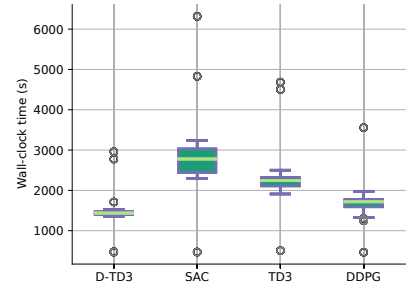}}
\hspace*{\fill}%
\caption{The algorithms' performance for the custom OpenAI gym continuous control task.}
\end{figure}
The difference between methods is that the proposed D-TD3 considers deterministic policy gradient and uses the return distribution learning approach while the re-tuned SAC method is based on a stochastic policy gradient and benefits from clipped double Q-learning. Besides, TD3 and DDPG are deterministic policy gradients where TD3 is based on a clipped double Q-learning. D-TD3 curve shows how agents learn to increase average reward and thereby decrease the costs of the network during the time concerning a large state space through interaction with different network configurations and parameters. The results reveal that
\begin{table}[!htb]
\caption{Average return over 3 trials of 2 million time steps. Maximum value for the task is bolded. $\pm$ corresponds to a single standard deviation over trials (runs).}
\scriptsize
\centering
\begin{tabular}{@{}lc@{}}\toprule
\textbf{Method} & \textbf{smartech-v1} \\ \midrule
\textbf{D-TD3} & \textbf{-0.11194465$\pm$0.0257}\\ \hdashline
\textbf{SAC}& -0.1327343$\pm$0.0240 \\ \hdashline
\textbf{TD3}& -0.13273136$\pm$0.0238 \\ \hdashline
\textbf{DDPG}&  -0.14393996$\pm$0.0227 \\ \hdashline

\bottomrule
\end{tabular}

\end{table}
traditional TD learning such as DDPG can suffer from overestimation and inconsistency in the form of propensity to converge to suboptimal policies throughout the learning, whereas clipped double Q-leaning algorithms can greatly reduce overestimation but they are prone to cause underestimation bias. Therefore, the return distribution learning has better performance for Q-value estimation accuracy and preventing gradient explosion that can be considered as an efficient method for continuous telecommunication control problems to attain SoA performance.\\
\emph{\textbf{-Time efficiency:}} As shown in figure 3-(b), we compare the time efficiency of different approaches. Results demonstrate that the D-TD3 method yields performance improvement in terms of the average wall-clock time consumption on the custom environment per 50 time steps and based on 100 evaluations. The time performance of D-TD3 is comparable to DDPG and much lower than TD3 and SAC. In fact, D-TD3 with a return distribution learning approach does not need actor-critic double Q-learning techniques and additional value or target networks to handle overestimations. All evaluations were run on a single computer with a 3.40 GHz 5 core Intel CPU.
\begin{figure*}[ht!]
\centering
\subfloat[Admission rate - Slice A]{%
      \includegraphics[width=0.24\textwidth]{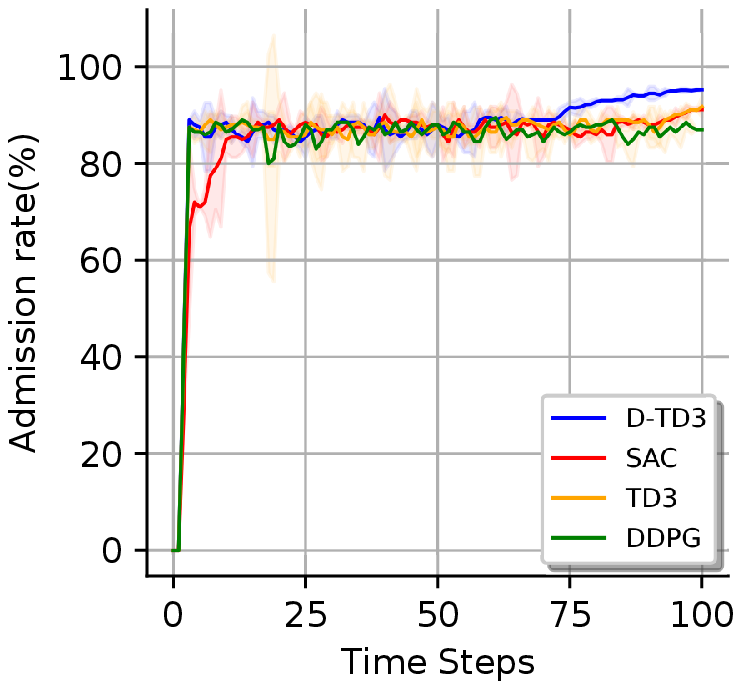}}
      \hfill
\subfloat[Admission rate - Slice B]{%
      \includegraphics[width=0.24\textwidth]{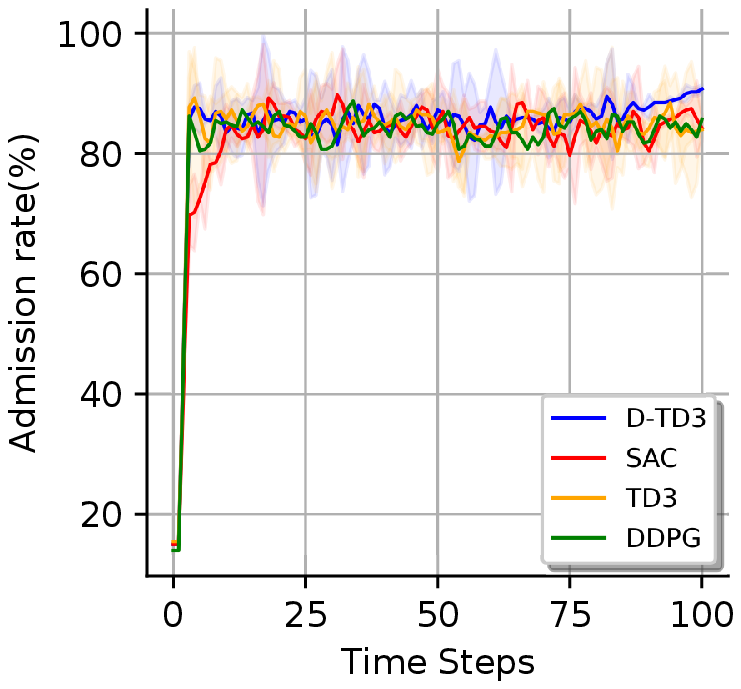}}
      \hfill
\subfloat[Admission rate - Slice C]{%
      \includegraphics[width=0.24\textwidth]{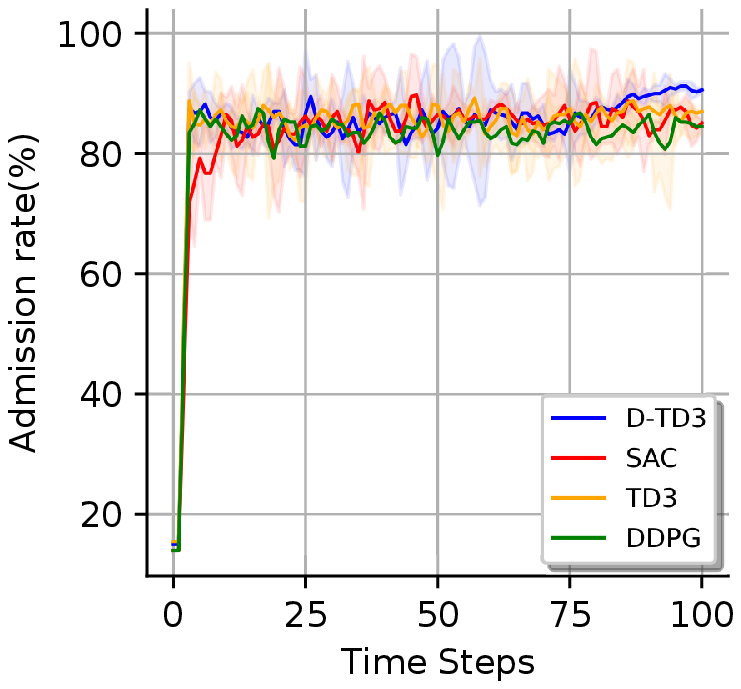}}
      \hfill
\subfloat[Latency - Slice A]{%
      \includegraphics[width=0.24\textwidth]{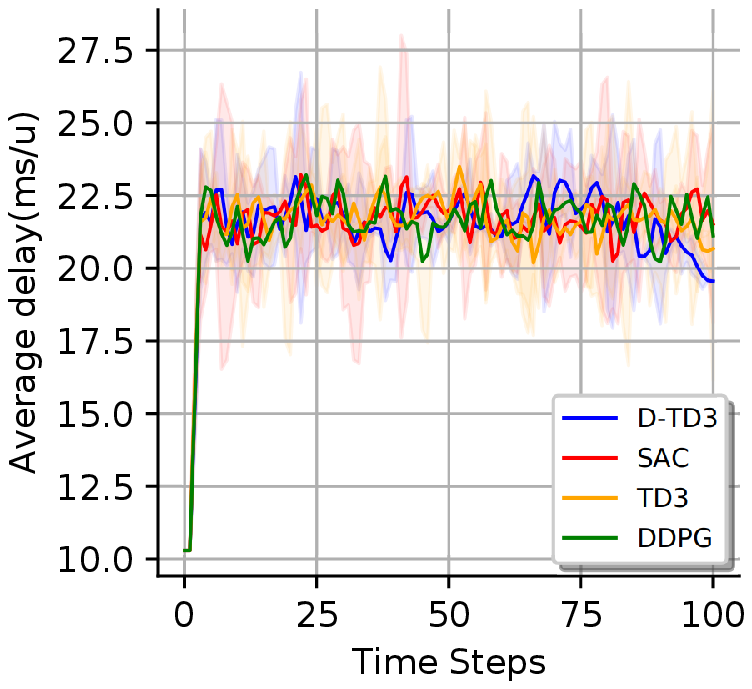}}
      \hfill
      \\
\subfloat[Latency - Slice B]{%
      \includegraphics[width=0.24\textwidth]{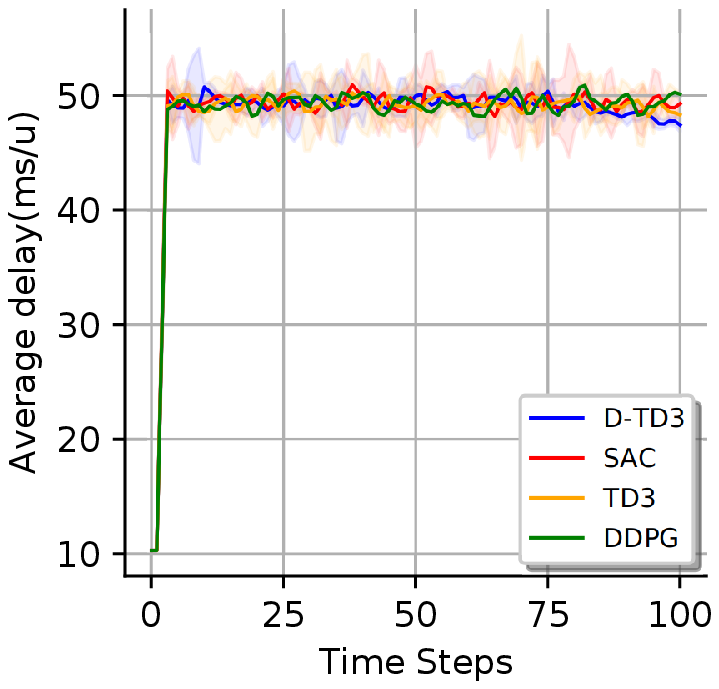}}
      \hfill
\subfloat[Latency - Slice C]{%
      \includegraphics[width=0.24\textwidth]{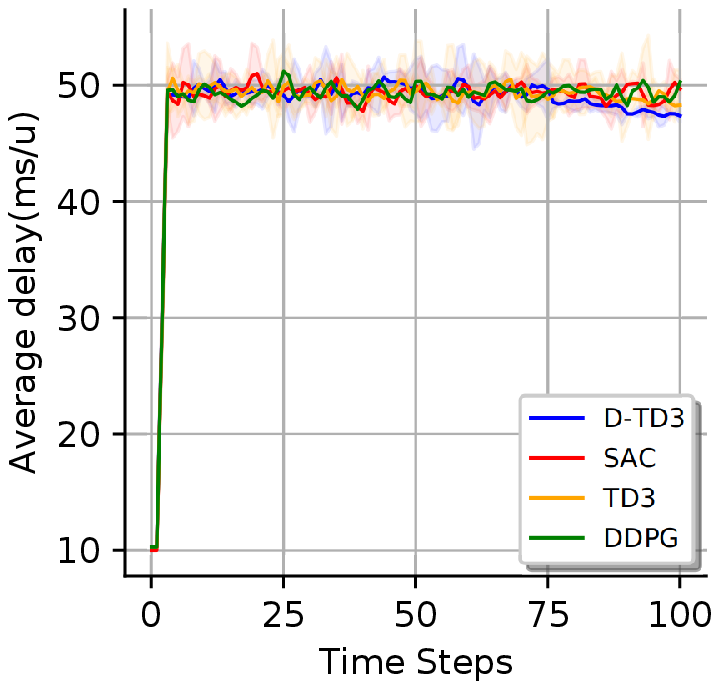}}
      \hfill
\subfloat[CPU utilization - Slice A]{%
      \includegraphics[width=0.24\textwidth]{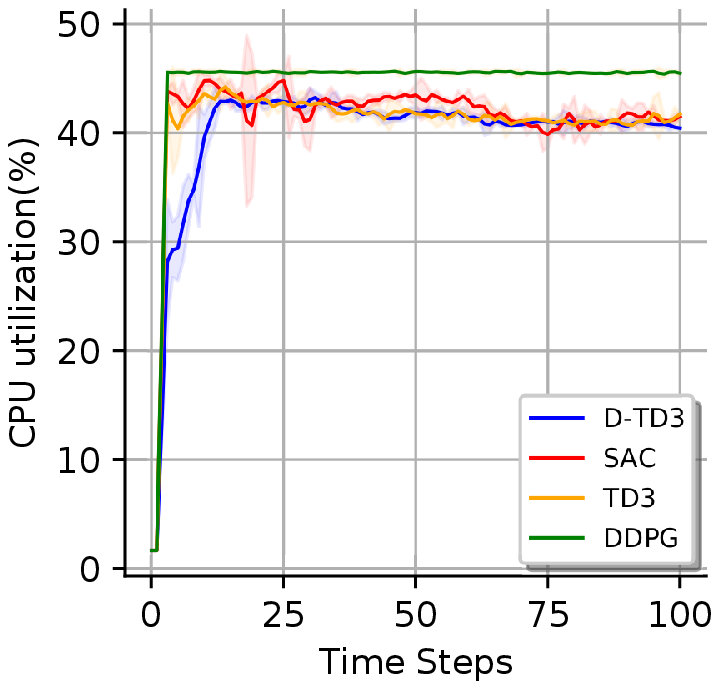}}
      \hfill
\subfloat[CPU utilization - Slice B]{%
      \includegraphics[width=0.24\textwidth]{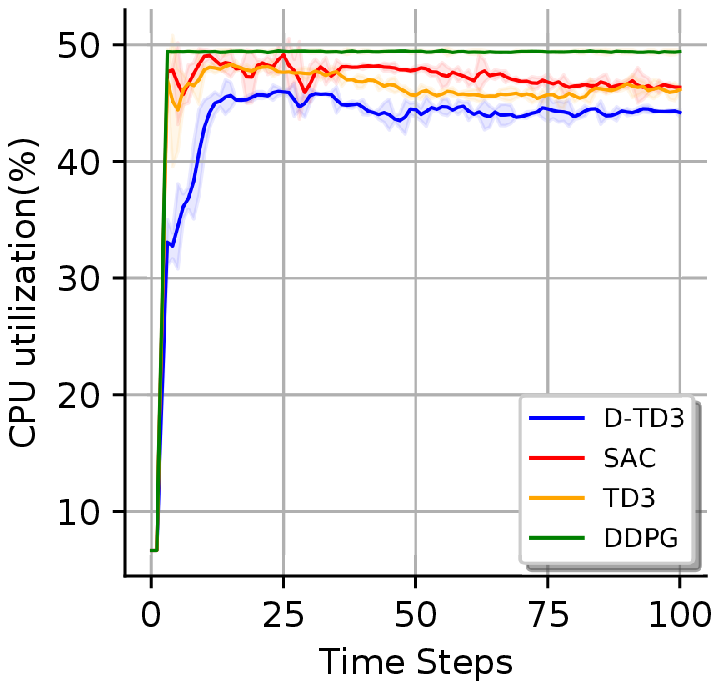}}\\
\subfloat[CPU utilization - Slice C]{%
      \includegraphics[width=0.24\textwidth]{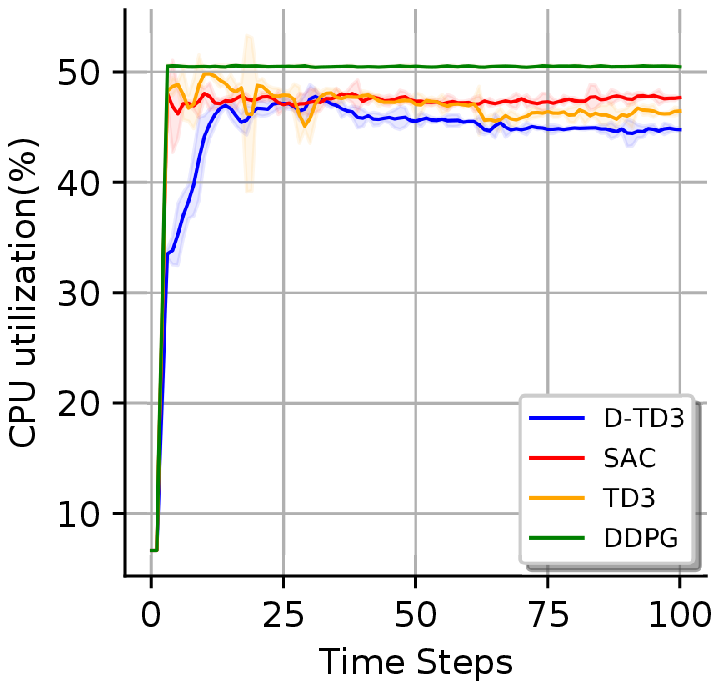}}
       \hfill
\subfloat[Energy consumption - Slice A]{%
      \includegraphics[width=0.24\textwidth]{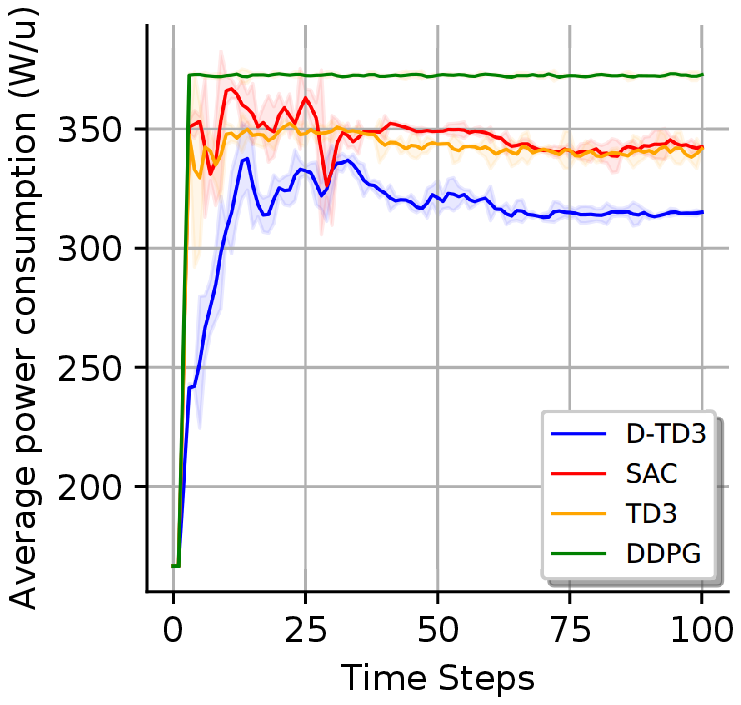}}
      \hfill
\subfloat[Energy consumption - Slice B]{%
      \includegraphics[width=0.24\textwidth]{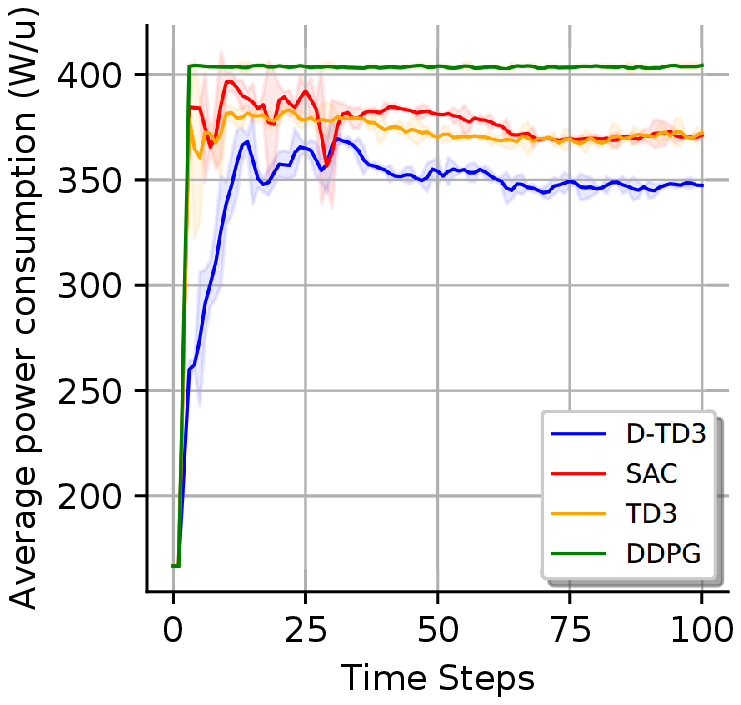}}
      \hfill
\subfloat[Energy consumption - Slice C]{%
      \includegraphics[width=0.24\textwidth]{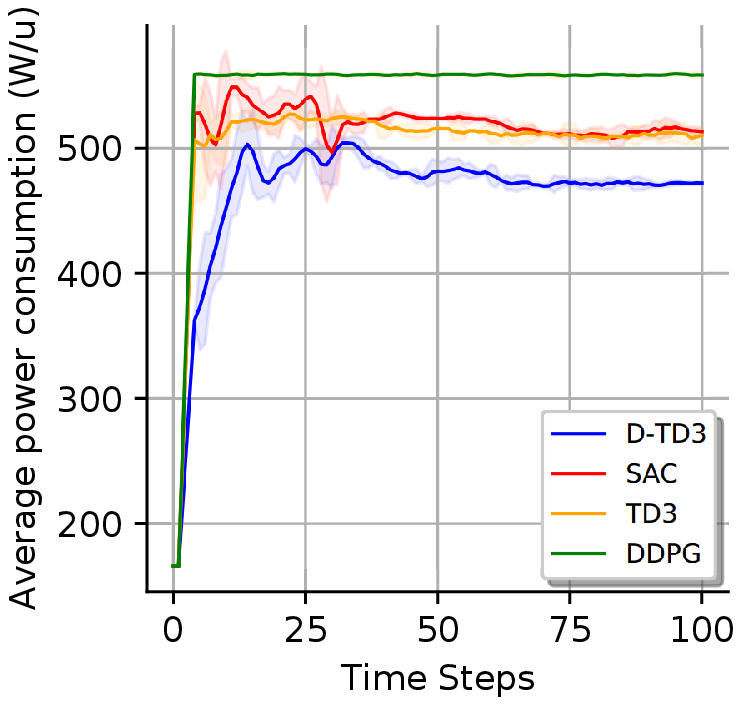}}

\caption{Network performance and costs comparison between D-TD3 and other DRL benchmarks. The curves are smoothed for visual clarity. The solid lines demonstrate the mean and the shaded regions correspond to confidence interval over 3 trials.}

\end{figure*}
\\
\emph{\textbf{-Admission Control:}} We define it as finding the best policy to maximize the admission rate of traffic or increase the number of users that can be admitted concurrently while the system optimally trades-off available resources to satisfy QoS requirements. It is essential in network slicing to control the admission of diverse requests belonging to different slices, especially the ultra-reliable low latency services. As shown in Figure 4-(a), 4-(b), and 4-(c), the trade-off strategy between MNO and tenants (slices) can mitigate resource contention and efficient solution in terms of financial costs and service availability. In essence, the D-TD3 algorithm learns to play the role of a demand-aware mechanism to update policies as long as new demands arrive in an abrupt sequence. The results show D-TD3 outperforms the other approaches concerning resource availability and constraints. The similarity between results can be interpreted according to the trading-off strategy between slices as results in high resource availability and high admission rate.  
\\
\emph{\textbf{-Latency:}} As explained in Sec. \rom{2}, latency originates from both the transmission in the access network and the processing in the central server. RAN slicing with cell-free architecture underpins latency-sensitive applications through shifting the computing tasks and VNFs to edge infrastructure located close to a large amount of APs and users. The estimation of both transmission and baseband processing values make our simulation environment realistic and taking into account the cross-layer factors that affect the latency. Figures 4-(d), 4-(e), and 4-(f) show how D-TD3 leads to less average delay per user compared to TD3, SAC, and DDPG. As mentioned before, agents maybe experience high fluctuations because of interaction with different configurations and large state-space to find the best action or policy in the network. Notice that slice A is more time-sensitive than other slices. 
\\
\emph{\textbf{-CPU utilization:}} To provide a novel vertical scaling policy, it is important to apply a mechanism for harvesting and freeing up the unused resources in network slicing according to demand prediction or demand-aware scheduling. A fine-grained resource control on enclosing services and between MNO and tenants has a direct impact on latency, energy, admission rate and also assuring service quality. We consider CPU utilization efficiency as a proportion of exploited CPU compared to total available CPU resources to execute a VNF. As depicted in Figures 4-(g), 4-(h), and 4-(i), D-TD3 deployment leads to more efficient usage of CPU compared to other algorithms. TD3 and the re-tuned SAC algorithms perform comparably and similarly at some time steps, but this can corroborate the significant role of clipped double Q-learning technique in actor-critic methods and solving network issues. As explained in Sec. \rom{4}, this technique takes the minimum value between the two independent estimates. We reduce the grater Q-estimate to the minimum value during the update to fend off overestimation and lead to safer policy updates. This value estimate approach is used as an approximate upper-bound to the true value estimate \cite{ref24}. Although it can induce an underestimation bias it is greatly preferable to overestimation because the value of underestimated actions does not tend to propagate during learning and policy updating \cite{ref26}-\cite{ref24}. In the implementation of the environment, we consider resource saturation issues where a central server is not completely saturated and reserve some part of resources for management software and also additional reserved resources in MNO for trading-off with critical services (i.e., Slice A).\\
\emph{\textbf{-Energy consumption:}} Another focal point in this paper is energy efficiency while pursuing a network slicing aware dynamic resource allocation approach as a new prominent figure of merit. Indeed, energy-efficient resource allocation is a key pillar in B5G networks. We consider the trade-off between CPU resource usage, latency, and energy consumption via defining a correlated multi-objective cost function. Figures 4-(j), 4-(k), and 4-(l) show that the performance of D-TD3 is better than other methods where the agent learns to satisfy other objectives and also minimize power consumption by decreasing VNFs instantiation and tuning wireless transmission power. As explained in Sec. \rom{4}, D-TD3 method benefits from curtailing and clipping technique to avoid gradient explosion and thereby stabilize model and learning procedure while obtaining the highest Q-value estimation and dramatically reduces overestimation. This objective is important because the predefined weight of this cost is more than other objectives. Indeed, the predefined weight can guide agents to know the importance and priority of a cost. As shown, the curves follow a very similar trend and the learning behavior is better than other objectives. Notice that a large part of power consumption in the network is constant and agents cannot change or improve these values and also in some scenarios, DDPG cannot learn perfectly and does not have stable learning behavior to solve a problem. \\
  \begin{figure}[ht!]
\centering
\hspace*{\fill}%
\subfloat[Admission rate]{%
      \includegraphics[width=0.23\textwidth]{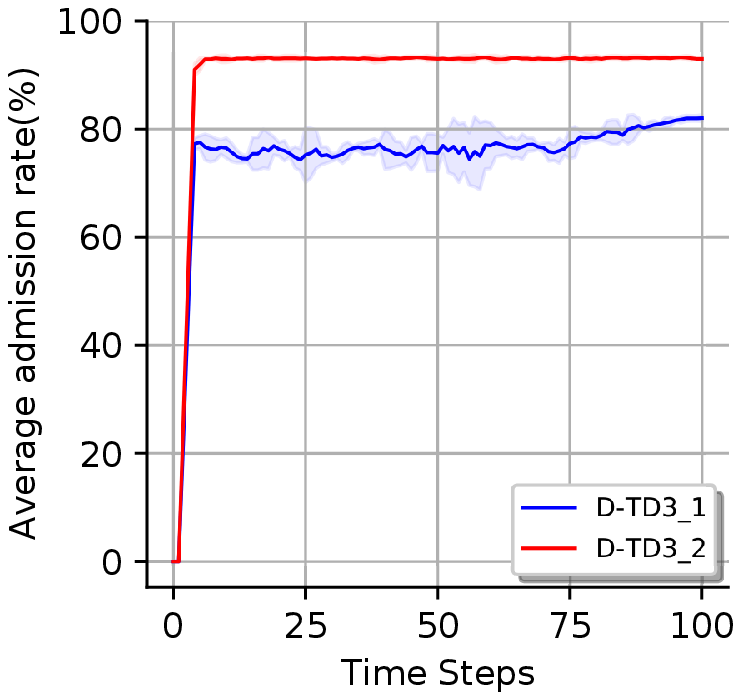}}
       \hfill
\subfloat[Network latency]{%
      \includegraphics[width=0.23\textwidth]{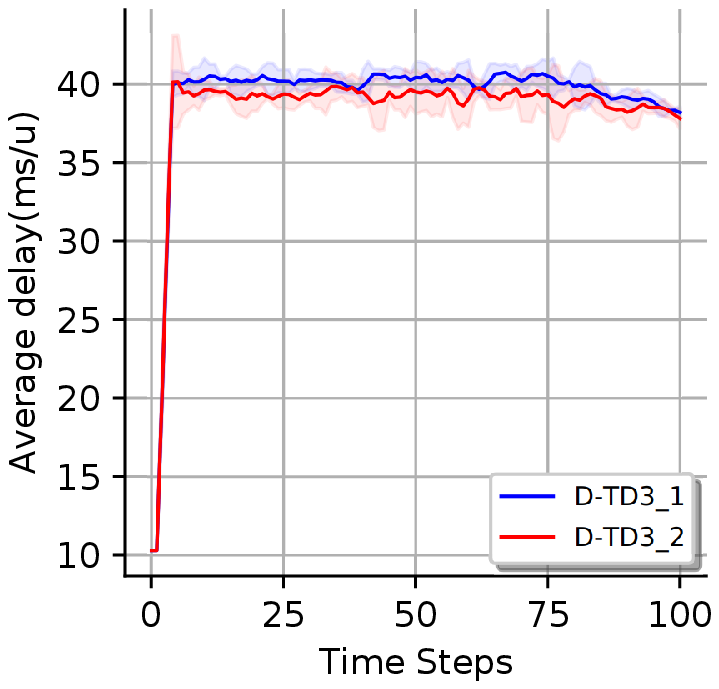}}
    \hfill
\subfloat[Energy consumption]{%
      \includegraphics[width=0.23\textwidth]{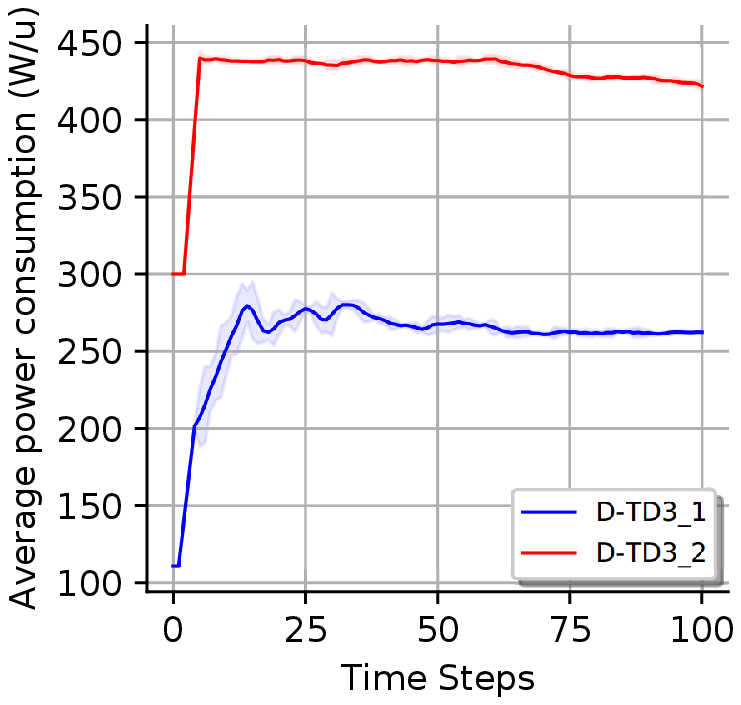}}
\hspace*{\fill}%

\caption{General network performance. The number of APs for blue curve is 150 and predefined weights $(\hat{\omega}_1, \hat{\omega}_2, \hat{\omega}_3, \hat{\omega}_4) = (1, 2, 1, 100)$ whereas we use 250 APs in simulation for red curve and set predefined weights to $(\hat{\omega}_1, \hat{\omega}_2, \hat{\omega}_3, \hat{\omega}_4) = (1, 1, 3, 100)$.}
\end{figure}
\emph{\textbf{-Performance of the entire network:}} We consider MNO and tenants as a unified network where slices are isolated and also have a trade-off with MNO. Figures 5-(a), 5-(b), and 5-(c) show the average performance of D-TD3 on the overall environment. The blue curve refers to predefined settings and parameters whereas the red curve shows the performance with different network parameters. As demonstrated, D-TD3 can have acceptable results based on different traffics and network states where it satisfies the objectives in terms of admission rate, latency, and energy consumption. Since the increase in the number of APs leads to better SINR and it can have a direct/indirect effect on cost functions and satisfying the constraints and thresholds in the network that reflect on increasing admission rate in figure 5-(a). The increase in the red curve is very slightly because we consider average on 3 slices in the second simulation. We increase priority and weight of latency to 3 as the main cost in our scenario. As shown in figure 5-(b) the red curve has a better performance compared to predefined parameters. It proves that the agent can distinguish between network costs through predefined weights. As depicted in figure 5-(c), the increase in the number of APs incurs more power consumption in the whole network. The red curve experiences a slight decline like the blue curve. Notice that the agent should pursue a multi-objective approach through trade-offs. For example, the increasing number of APs improves admission rate while incurs more power consumption, and thereby we cannot expect a huge improvement in all metrics. 
\section{Conclusion}
In this paper, we have proposed a lifelong zero-touch network slicing approach and developed the resource allocation policy with well-founded and cross-layer models. In contrast to existing works on AI and telecommunication, we have defined and corroborated via extensive experimental results to show how AI can help to find a solution to minimize the correlated multi-objective cost function. We have investigated the potential of applying a continuous DRL method called D-TD3 and proposed techniques on resource management tasks in network slicing which outperforms the other SoA policy-based and continuous algorithms. In this intent, we have built a B5G RAN environment that includes a slice-enabling cell-free mMIMO access architecture using OpenAI Gym toolkit where, thanks to its standardized interface, it can be easily tested with different DRL schemes and enables the central server to learn how re-configure computing resources autonomously, aiming at minimizing the latency, energy consumption and VNF instantiation for each slice and MNO. In our actor-critic method, we have leveraged a state-action return distribution learning approach and proposed a replay policy based on a prioritized asynchronous actor-learner model with reward-penalty techniques to speed-up training and stabilize the learning process in network slicing environment.

As future research directions, we consider addressing the E2E massive slices. The problem addressed in this work does not consider the variety in type of VNFs and also VNFs chains. Moreover, we should consider more resources such as PRBs and memory requirements. This factor highly increases the complexity of admission control and resource allocation problem. Besides, another interesting direction might be the comparison of our method with different stochastic policy gradient approaches in actor-critic methods.


\balance
\end{document}